\documentclass[aps, prb, preprint, showpacs, superscriptaddress, floatfix,amsmath,amssymb,longbibliography]{revtex4-1}
\usepackage[english]{babel}
\usepackage{bm}
\usepackage{graphicx}
\usepackage{subfigure}
\usepackage{appendix}
\usepackage{dcolumn}
\usepackage{soul} 
\usepackage[nottoc]{tocbibind}
\usepackage{dcolumn}
\usepackage{hhline}

\usepackage{color}
\newcommand{\bra}[1]{\langle #1|}
\newcommand{\ket}[1]{|#1\rangle}

\begin{document}

\title{Multi-valley envelope function equations and effective potentials for P impurity in silicon}

\author{M.V. Klymenko}
\affiliation{Department of Chemistry, B6c, University of Liege, B4000 Liege, Belgium}

\author{S. Rogge}
\affiliation{School of Physics, The University of New South Wales, Sydney, New South Wales 2052, Australia}

\author{F. Remacle}
\email{fremacle@ulg.ac.be}
\affiliation{Department of Chemistry, B6c, University of Liege, B4000 Liege, Belgium}

\pacs{61.72.uf, 73.22.-f, 71.55.-i, 71.55.Ak, 71.18.+y}

%%%%%%%%%%%%%%%%%%%%%%%%%%%%%%%%%%%%%%%%%%%%%%%%%%%%%%%%%%%%%%%%%%%%%%%%%%%%%%%%%%%%%%%%%%%%

\begin{abstract}
    We propose a system of real-space envelope function equations without fitting parameters for modeling the electronic spectrum and wave functions of a phosphorus donor atom embedded in silicon. The approach relies on the Burt-Foreman envelope function representation and leads to coupled effective-mass Schroedinger equations containing smooth effective potentials. These potentials result from the spatial filtering imposed on the exact potential energy matrix elements in the envelope function representation. The corresponding filter function is determined from the definition of the envelope function. The resulting effective potentials and the system of envelope functions jointly reproduce the valley-orbit coupling effect in the doped silicon. Including the valley-orbit coupling not only of the 1s, but also for 2s atomic orbitals, as well as static dielectric screening is found crucial to accurately reproduce experimental data. The measured binding energies are recovered with a maximum relative error of 1.53 \%. The computed wave functions are in a good agreement with experimental measurements of the electron density provided by scanning tunneling microscopy. 
\end{abstract}

\maketitle

\section{Introduction}

Recent technological advances in deterministic doping \cite{SAT} get closer to implementations of devices performing classical or quantum computations on a single donor atom \cite{SAT_book} as has been suggested by Kane \cite{Kane}. In parallel, significant progress has been made in scanning tunnelling microscopy (STM) imaging of impurity atoms embedded in silicon several nanometers below the surface, in their ground and excited states \cite{SvenVO,SvenSTM}. Knowing the electronic structure of impurity atoms is essential for understanding the physics of device operation. Moreover, the interpretation of the measured STM images requires the accurate modeling not only of the energy spectra, but also of the real-space wave functions of the impurities. The same is true for analysing the hyperfine structure of the energy spectra of shallow donors where the charge density at the impurity nucleus determines the energy splitting \cite{hf,Pica}.

Computational approaches for determining the electronic structure of impurities in silicon can be divided into two classes: supercell methods and effective-mass techniques. The tight-binding method  \cite{TB} and methods based on density functional theory with pseudopotentials and plane-wave expansions belong to the first class. Since the effective Bohr radius \cite{Wellard1} of a phosphorus donor atom in silicon is 3.1 nm, the size of the supercell used in those methods varies from $10^5$ to $10^7$ atoms \cite{TB}. These approaches can be implemented to any required numerical accuracy but remain very demanding for computational resources.

In effective mass theory, all relevant information about the band structure of bulk silicon is contained in a small number of so-called band structure parameters. However, applying successfully effective mass theory for donor atoms in silicon is challenging due to the strong central cell attractive potential that leads to valley-orbit coupling  \cite{Baldereschi}.

The pioneering work of Pantelides and Sah \cite{Pantelides} led to significant progress in the modeling the spherically-symmetric central-cell part of the impurity potential by using \textit{ab initio} computations. In Ref. [\onlinecite{Pantelides}], Pantelides and Sah have demonstrated the important role of the static dielectric screening in the central-cell potential that has been also justified by recent tight-binding computations \cite{Sven_potential}. Usually, the expected tetrahedral symmetry, which is responsible for valley-orbit splitting, is imposed not on the potential itself, but on the wave function using group-theoretical considerations like in the Friztsche-Twose equations \cite{Twose}. This approach works well for non-interacting impurities in bulk silicon, but its extension to impurity clusters or any non-spherically-symmetric confinement potentials is not trivial. Moreover while the energy spectrum is reproduced accurately, the corresponding wave functions do not fit to the current experimental data from the STM measurements \cite{SvenSTM}. Thus, the central-cell part of the phosphorus donor potential remains the subject of intense studies \cite{Overhof,Castner,Greenman,Gamble}. Recent \textit{ab initio} calculations evidence a tetrahedral symmetry of the central-cell potential of the donor impurity  \cite{Castner,Greenman,Gamble} as well as a small displacement of its silicon neighbouring atoms \cite{Overhof}. It has been also established in Refs. [\onlinecite{Wellard}] and [\onlinecite{Abinitio}] that computing accurately the valley-orbit splitting requires detailed information on the periodic Bloch functions which has to be computed beyond the effective mass approximation. In modern effective-mass methods, the energy spectrum is modeled using a set of fitting parameters and ad-hoc corrections aimed to reproduce valley-orbit splitting of the ground state energy level and/or electron charge density at the impurity nucleus \cite{Wellard,Pica,Gamble}. The number of fitting parameters varies from a single one in [\onlinecite{Wellard}] to five in [\onlinecite{Gamble}] (most often three fitting parameters are used \cite{Friesen,us}). In the paper of Gamble et al. \cite{Gamble}, the proper tetrahedral symmetry has been imposed directly on the central-cell potential. Although effective-mass approaches with fitting parameters can reproduce energy spectra and electron densities at the impurity nuclei accurately, the overall shape of the wave functions is not guaranteed to be correct. The shape and localization of the wave function are crucial to compute accurate electron-electron correlation effects in systems consisting of several interacting donors \cite{Abinitio, Wellard, Exchange}.

The goal of this paper is to develop a real-space effective mass approach without fitting parameters using a small number of systematically controllable approximations. In the framework of these approximations, the approach should guarantee the correct overall shape of the wave function and reproduce the energy spectrum. To avoid using fitting parameters, instead of the conventional envelope-function approximation  \cite{Luttinger-Kohn,wave_mech}, we derive the system of the envelope function equations starting from the exact Burt-Foreman envelope function representation. We show that this representation is equivalent to linear combination of bulk bands or plane-wave expansions. Using an exact envelope-function representation enables to implement systematic series truncation and to associate each envelope function with a defined region in the k-space. 

The resulting envelope function equations are formally equivalent to the Shindo-Nara equations \cite{Shindo-Nara} and contain a smooth effective potential which is derived  \textit{ab initio} and depends on periodic Bloch functions and on the attractive potential of the impurity. The effective potential results from a low-pass filtering procedure imposed by the periodic boundary conditions of the crystal lattice and from a systematic truncation of approximating series (e.g. the single-band approximation). We compute the effective potential using the point-charge potential
 with a static screening as a model for the central-cell potential of the impurity atom \cite{Pantelides}. This model is valid for P. For non-isocoric impurities, such as As, the point charge must be replaced by a distance-dependent potential that can be computed using density functional theory or the \textit{ab initio} technique described in [\onlinecite{Pantelides}].

The crystal symmetry is introduced in the basis set by using the periodic Bloch functions computed at the level of the density functional theory with the local density approximation (DFT-LDA). Models combining the effective mass method and \textit{ab initio} computations have been used before by several authors  \cite{Wellard,Abinitio,PhysRevB.84.155320}. Our model also relies on that approach, however it is implemented so that all atomistic details of the wave function within the unit cell are mapped to smooth real-space potentials without any fitting with experimental data. These potentials can be viewed as local pseudo-potentials and further used in effective mass calculations.

The computational method should take into account complicated configurations of external electrostatic fields and confinement potential. In the effective mass approach, this can be implemented variationally either by using predefined basis sets with several variational parameters which can be adjusted to electrostatic fields \cite{Debernardi} or by using a grid method in real or momentum space. Here, we compute the wave function using a combined method: the Schroedinger equation is first solved neglecting the valley-orbit coupling for an arbitrary confinement potential and external electrostatic fields by a real space grid method, and, then the computed wave functions are used as a basis set in a variational procedure that diagonalizes the Schroedinger equation that includes the valley-orbit coupling. This makes our approach flexible and well adapted for any silicon nanostructures.

The paper is organized as follows: in Section II we define the multi-valley envelope function and show its relations with the plane-wave expansion and the linear-combination of bulk bands representation. We then derive the envelope function equation. In Section III, we derive the expressions for the potential energy terms and compute them using a screened Coulomb potential with a static screening and periodic Bloch functions from DFT-LDA calculations. In Section IV we present our numerical approach for solving the system of envelope function equations and show results of energy spectrum computations for P donor atom in silicon. We then provide an analysis of the donor wave functions. Concluding remarks are given in Section V.

\section{Multi-valley envelope function}

\subsection{Multi-valley envelope functions representation: definition}

The wave function for structures with a periodic potential may be expanded in plane waves (PW) as follows:

\begin{equation}
    \psi(\mathbf{r})=\sum\limits_{\mathbf{G},\mathbf{k}_0} \sum\limits_{\mathbf{k}\in SBZ_{\mathbf{k}_0}}
    \tilde{\psi}_{\mathbf{G},\mathbf{k}_0+\mathbf{k}}e^{i(\mathbf{G}+\mathbf{k}_0+\mathbf{k})\mathbf{r}}, 
    \label{pw}
\end{equation}
where $\tilde{\psi}_{\mathbf{G},\mathbf{k}_0+\mathbf{k}}$ are Fourier coefficients, $\mathbf{G}$ are the reciprocal lattice vectors and
$\mathbf{k}_0+\mathbf{k}$ is a wave vector within the first Brillioun zone.

In the expansion (\ref{pw}) each wave vector within the first Brillouin zone (BZ) is determined as the unique sum of two vectors, $\mathbf{k}_0+\mathbf{k}$,
where $\mathbf{k}_0$ specifies a region inside the BZ and the wave vector,
$\mathbf{k}$, is bound inside that region. Such a partitioning of the BZ may be done in different ways depending on the specific problem considered. For silicon, it is convenient to consider six regions: each represents a sector of the BZ (SBZ) such that it contains a single conduction band valley (see Fig. 1). In this case, the vector $\mathbf{k}_0$ points to one of six conduction band minima \cite{us}.

The plane waves, $e^{i\mathbf{G}\mathbf{r}}$, in Eq. (1),
have the periodicity of the crystal lattice. Following the methodology proposed by Burt 
\cite{Burt,Burt3},each of them can be expanded in terms of periodic Bloch functions, which form a complete basis set of periodic functions for each point of the BZ. The usual practice in $k\cdot p$ theory \cite{Voon} is to use the basis set,
 $u_{n,\mathbf{k}}(\mathbf{r})$, taken from a single point in the BZ (the most common case is to use the center of the BZ), we are free to chose any reference point for different regions. Specifically, in each region, we expand $e^{i\mathbf{G}\mathbf{r}}$ in terms of the periodic Bloch functions, $u_{n,\mathbf{k}_0}(\mathbf{r})$, taken at the wave vector corresponding to the conduction band minimum. As a result, Eq. (\ref{pw}) reads:
 
\begin{figure}[t]
\centering
\includegraphics[width=7.5 cm]{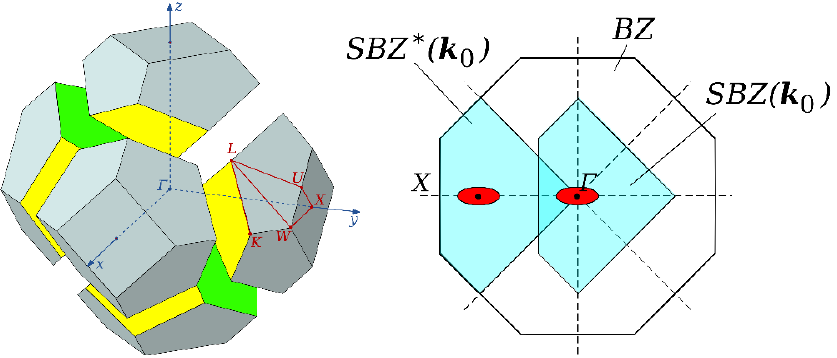}
\caption{Partitioning of the Brillouin zone (BZ) into six sectors (SBZ), each of which is related to a multi-valley envelope function defined for a specific valley of the conduction band.} 
\label{BZ}
\end{figure}

\begin{equation}
    \psi(\mathbf{r})=\sum\limits_{n,\mathbf{G},\mathbf{k}_0} \sum\limits_{\mathbf{k}\in SBZ_{\mathbf{k}_0}}
    \tilde{\psi}_{\mathbf{G},\mathbf{k}_0,\mathbf{k}}\tilde{u}_{n,\mathbf{G}}(\mathbf{k}_0) u_{n,\mathbf{k}_0}(\mathbf{r})e^{i(\mathbf{k}_0+\mathbf{k})\mathbf{r}}, 
    \label{pw1}
\end{equation}
where  $\tilde{u}_{n,\mathbf{G}}(\mathbf{k}_0)$ is the PW expansion coefficients for periodic Bloch functions.

Defining the multi-valley envelope function as:

\begin{equation}
    f_{n,\mathbf{k}_0}(\mathbf{r}) = \sum\limits_{\mathbf{G}} \sum\limits_{\mathbf{k}\in SBZ_{\mathbf{k}_0}} \tilde{\psi}_{\mathbf{G},\mathbf{k}_0,\mathbf{k}}\tilde{u}_{n,\mathbf{G}}(\mathbf{k}_0)e^{i \mathbf{k}\mathbf{r}}, 
    \label{ef}
\end{equation}

the multi-valley envelope function representation of the wave function given in Eq. (\ref{pw}) becomes:

\begin{equation}
    \psi(\mathbf{r}) = \sum\limits_{n,\mathbf{k}_0} f_{n,\mathbf{k}_0}(\mathbf{r}) u_{n,\mathbf{k}_0}(\mathbf{r})e^{i \mathbf{k}_0 \mathbf{r}},
    \label{wf}
\end{equation}

The envelope function in Eq. (\ref{wf}) is an exact and unique representation
\cite{Burt, Foreman_mix} of the wave function developed by Burt and Foreman.
Formally the Burt-Foreman and Luttinger-Kohn envelope functions are equivalent at the level of definitions. Both of them obey the important feature that their  Fourier components lie strictly inside the first BZ by definition (in our case inside the SBZ). Unlike in the envelope function approximation proposed by Luttinger and Kohn \cite{Luttinger-Kohn}, in the Burt-Foreman envelope function representation this feature is preserved in all derivations of the envelope-function equations and defines the smoothness and non-local properties of the effective potential entering into those equations. The Burt-Foreman envelope function always includes explicitly a band index allowing for the band mixing effect. We have modified the original definition of the envelope function by adding the valley index $\mathbf{k}_0$ allowing for the valley-orbit mixing.

For an infinite number of regions, the expansion (\ref{wf}) tends to the full-Brillouin zone approach \cite{Zunger, Bertazzi}. A larger number of regions enhances the accuracy of the numerical solution, at the expense, however, of the number of equations to solve.

\subsection{Relationships between different representations}

The electronic structure of a single impurity atom in the crystal lattice can be computed in several representations. The most commonly used are plane wave (PW) expansion, envelope function (EF) representation  \cite{Burt3} and linear combination of bulk
bands (LCBB) \cite{Zunger}. They are summarized in Eq. (\ref{rep}):

\begin{equation}
    \psi(\mathbf{r})=\begin{cases}
        \sum\limits_{n,\mathbf{k}_0} f_{n,\mathbf{k}_0}(\mathbf{r})
        u_{n,\mathbf{k}_0}(\mathbf{r})e^{i \mathbf{k}_0 \mathbf{r}}, &
        \text{EF};\\
        \sum\limits_{\mathbf{G},\mathbf{k}_0,\mathbf{k}}
        \tilde{\psi}_{\mathbf{G},\mathbf{k}_0+\mathbf{k}}e^{i(\mathbf{G}+\mathbf{k}_0+\mathbf{k})\mathbf{r}}, &
        \text{PW};\\
        \sum\limits_{n,\mathbf{k}_0,\mathbf{k}}
        c_{n,\mathbf{k}_0+\mathbf{k}}u_{n,\mathbf{k}_0+\mathbf{k}}(\mathbf{r})e^{i(\mathbf{k}_0+\mathbf{k})\mathbf{r}}, &
        \text{LCBB}.
    \end{cases}
    \label{rep}
\end{equation}

Each representation leads to correct results. Their computational efficiency depends on the problem to which they are applied. For example, the EF method is most efficient when the confinement potential varies slowly in real space. LCBB is most convenient when a specific mixing of electronic states, like  $\Gamma$-$X$ valley mixing in GaAs/AlAs quantum dots \cite{Zunger1}, is known \textit{a priori} that allows to reduce the size of the basis set formed from bulk states taken over the whole BZ. The PW expansion works obviously very well for periodic structures such as semiconductor superlattices  \cite{wave_mech}.

The representations mentioned above are related via unitary transformations. We derive explicitly the corresponding unitary matrices since they are important for the developments below. The unitary matrices are obtained by Fourier transforms of all coordinate-dependent factors in Eqs.  (\ref{rep}):

\begin{equation}
    \psi(\mathbf{G},\mathbf{k}_0,\mathbf{k})= \begin{cases}
        \sum\limits_{n} u_{\mathbf{G},n}(\mathbf{k}_0) \tilde{f}_{n}(\mathbf{k}_0,\mathbf{k})
        , & \text{EF};\\
        \tilde{\psi}_{\mathbf{G},\mathbf{k}_0+\mathbf{k}}, &
        \text{PW};\\
        \sum\limits_{n} u_{\mathbf{G},n}(\mathbf{k})c_{n,\mathbf{k}_0+\mathbf{k}}, &
        \text{LCBB}.
    \end{cases}
    \label{rep_f}
\end{equation}

In Eqs. (\ref{rep_f}) the sums over band indices $n$ can be considered as a matrix multiplication by treating $u_{\mathbf{G},n}(\mathbf{k}_0)$ as an element of a square matrix $U(\mathbf{k}_0)$ parametrically dependent on $\mathbf{k}_0$
with indices $\mathbf{G}$ and $n$. Correspondingly,
$\tilde{f}_{n}(\mathbf{k}_0,\mathbf{k})$,
$\tilde{\psi}_{\mathbf{G},\mathbf{k}_0+\mathbf{k}}$ and
$c_{n,\mathbf{k}_0+\mathbf{k}}$ may be
considered as elements
of vectors  $F(\mathbf{k}_0,\mathbf{k})$, $\Psi(\mathbf{k}_0,\mathbf{k})$ and
$C(\mathbf{k}_0,\mathbf{k})$. Taking into account that the matrices $U(\mathbf{k})$
are unitary \cite{Voon} for all wave vectors
$\mathbf{k}$ we readily recover the relations between the three representations of Eq. (\ref{rep}) in matrix form:

\begin{equation}
    \begin{cases}
        \Psi(\mathbf{k}_0,\mathbf{k})=U(\mathbf{k})C(\mathbf{k}_0,\mathbf{k}),\\
        \Psi(\mathbf{k}_0,\mathbf{k})=U(\mathbf{k}_0)F(\mathbf{k}_0,\mathbf{k}),\\
        C(\mathbf{k}_0,\mathbf{k})=U^{\dagger}(\mathbf{k})U(\mathbf{k}_0)F(\mathbf{k}_0,\mathbf{k}). 
    \end{cases}
    \label{can}
\end{equation}

\subsection{Envelope function equations and $k\cdot p$ method}

Our goal is to derive real-space differential equations, the solutions of which define the multi-valley envelope functions in periodic media with known band structure, $E(\mathbf{k})$, in the presence of an additional non-periodic potential
$V(\mathbf{r})$. We start with
the equation for LCBB \cite{Zunger} in matrix form in momentum space:

\begin{equation}
    E(\mathbf{k}_0+\mathbf{k})C(\mathbf{k})+\sum\limits_{\mathbf{k}'_0,\mathbf{k}'}
    V_{LCBB}(\mathbf{k}_0+\mathbf{k},\mathbf{k}'_0+\mathbf{k}')C(\mathbf{k}')=\varepsilon C(\mathbf{k}), 
    \label{LCBB}
\end{equation}
where $E(\mathbf{k}_0+\mathbf{k})$ is the diagonal matrix containing the set of band energies for the wave vector $\mathbf{k}_0+\mathbf{k}$ and an element of the matrix representation of the non-periodic potential reads $\left[V_{LCBB}(\mathbf{k}_0+\mathbf{k},\mathbf{k}'_0+\mathbf{k}')\right]_{nm}=\bra{n,\mathbf{k}_0+\mathbf{k}}V(\mathbf{r})\ket{m,\mathbf{k}'_0+\mathbf{k}'}$. 

Using the canonical transformation (\ref{can}), Eq. (\ref{LCBB}) can be rewritten in the envelope function representation:
\begin{widetext}
\begin{equation}
    U^{\dagger}(\mathbf{k}_0)
    H(\mathbf{k}_0+\mathbf{k})U(\mathbf{k}_0)F(\mathbf{k}_0,\mathbf{k})
    +\sum\limits_{\mathbf{k}'_0,\mathbf{k}'}U^{\dagger}(\mathbf{k}_0)
    V_{PW}(\mathbf{k}_0+\mathbf{k},\mathbf{k}'_0+\mathbf{k}')U(\mathbf{k}_0') 
    F(\mathbf{k}_0',\mathbf{k}') =\varepsilon F(\mathbf{k}_0,\mathbf{k}), 
    \label{large}
\end{equation}
\end{widetext}
where
$\left[V_{PW}(\mathbf{k}_0+\mathbf{k},\mathbf{k}'_0+\mathbf{k}')\right]_{\mathbf{G},\mathbf{G}'}=\bra{\mathbf{G}+\mathbf{k}_0+\mathbf{k}}V(\mathbf{r})\ket{\mathbf{G}'+\mathbf{k}'_0+\mathbf{k}'}$
is an element of the plane wave matrix representation of the non-periodic potential. The potentials in Eq. (\ref{LCBB}) and Eq. (\ref{large}) are related to each other by
$V_{LCBB}(\mathbf{k}_0+\mathbf{k},\mathbf{k}'_0+\mathbf{k}')=U^{\dagger}(\mathbf{k})
V_{PW}(\mathbf{k}_0+\mathbf{k},\mathbf{k}'_0+\mathbf{k}')U(\mathbf{k}')$. The matrix
$H(\mathbf{k}_0+\mathbf{k})=U(\mathbf{k})E(\mathbf{k}_0+\mathbf{k})U^{\dagger}(\mathbf{k})$ is the PW representation of the periodic part of the Hamiltonian, its matrix elements read:

\begin{equation}
    \left[H(\mathbf{k}_0+\mathbf{k})
        \right]_{\mathbf{G},\mathbf{G}'}=T_{\mathbf{G}}(\mathbf{k}_0,\mathbf{k})\delta_{\mathbf{G},\mathbf{G}'}+V_{\mathbf{G},\mathbf{G}'},
\end{equation}
where
$$T_{\mathbf{G}}(\mathbf{k}_0,\mathbf{k})=\left[\frac{|\mathbf{G}|^2}{2}+i\left((\mathbf{k}_0+\mathbf{k}) \cdot \mathbf{G}\right)+
        \frac{(\mathbf{k}_0+\mathbf{k})^2}{2} \right],$$
and
$V_{\mathbf{G},\mathbf{G}'}$ is the plane wave representation of the periodic crystal potential.

Eq. (\ref{large}) is general and it is exact since no approximation has been made up to now. In  Eq. (\ref{large}), both terms on the left-hand side (the periodic and non-periodic ones) are non-diagonal matrices responsible for band mixing. The problem can be partially simplified by the diagonalizing periodic part of the Hamiltonian with the k$\cdot$p method.
This technique is based on canonical transformations and second-order pertubation theory. It allows for certain bands belonging to a set A to take into account the interband mixing with all others bands of a set B. It leads to the effective mass Hamiltonian for the set A: $\left[ U^{\dagger}(\mathbf{k}_0)
H(\mathbf{k}_0+\mathbf{k})U(\mathbf{k}_0)\right]_{nm} \rightarrow \left[
H_{kp}\left(\mathbf{k}_0, \mathbf{k}  \right)\right]_{nm} \delta_{n \in A} \delta_{m \in
A}$. The energy states of donor atoms in silicon lie close to conduction bands in the band gap. Therefore we restrict the set A to the lowest conduction band. A very common approximation made at this stage is that the non-periodic potential does not lead to band mixing between sets A and B so the canonical transformations do not affect the non-periodic potential (see Ref. [\onlinecite{Luttinger-Kohn}]). Since we do not intend to use fitting parameters in this work, we take this approximation as an ansatz and will check its validity by comparing our computed results with experimental values. After applying the k$\cdot$p method, the envelope function equations read:

\begin{eqnarray}
    H_{kp}&&\left(\mathbf{k}_0, \mathbf{k} \right) F(\mathbf{k}_0,\mathbf{k}) \nonumber \\
    &&+\sum\limits_{\mathbf{k}_0',\mathbf{k}'}U^{\dagger}(\mathbf{k}_0)
    V_{PW}(\mathbf{k}_0+\mathbf{k},\mathbf{k}'_0+\mathbf{k}')U(\mathbf{k}_0') 
    F(\mathbf{k}_0',\mathbf{k}') \nonumber \\
    &&=\varepsilon F(\mathbf{k}_0,\mathbf{k}),
    \label{kp}
\end{eqnarray}
where $H_{kp}\left(\mathbf{k}_0, \mathbf{k} \right)$ is the single-band k$\cdot$p-Hamiltonian for bulk silicon. The Hamiltonian $H_{kp}\left(\mathbf{k}_0, \mathbf{k} \right)$ for silicon is known \cite{Luttinger-Kohn}, so we will further pay attention to the potential energy term.

\section{Potential energy term}

\subsection{General real-space expression}

In element-wise form the potential energy term in Eq. (\ref{kp}) reads: 
\begin{widetext}
\begin{equation}
    Vf=
    \sum\limits_{m,\mathbf{k}'}\sum\limits_{\mathbf{G},\mathbf{G}'}u_{n,\mathbf{G}}^{*}(\mathbf{k}_0)
    \tilde{v}(|\mathbf{G}'-\mathbf{G}+\mathbf{k}'_0+\mathbf{k}'-\mathbf{k}_0-\mathbf{k}|)u_{\mathbf{G}',m}(\mathbf{k}_0') 
    \tilde{f}_m(\mathbf{k}_0',\mathbf{k}'),
\end{equation}
\end{widetext}
where $\tilde{v}(|\mathbf{G}'-\mathbf{G}+\mathbf{k}'_0+\mathbf{k}'-\mathbf{k}_0-\mathbf{k}|)$ are the PW expansion coefficients for the impurity potential.

Acting with the linear operator $\frac{1}{L^3} \sum\limits_{\mathbf{k}}e^{i\mathbf{k}\mathbf{r}} \times$ from the left-hand side, each term of Eq. (\ref{kp}) can be transformed into real space. Particularly, the potential term becomes:
\begin{widetext}
\begin{equation}
    Vf=\sum\limits_{m,\mathbf{k}'_0} \int d\mathbf{r}' f_m (\mathbf{k}_0',\mathbf{r}') \int d\mathbf{r}'' u_{n}^{*}(\mathbf{k}_0,\mathbf{r}'')
    V(\mathbf{r}'')u_{m}(\mathbf{k}_0',\mathbf{r}'') \Delta_{\mathbf{k}_0}(\mathbf{r} - \mathbf{r}'')
    \Delta_{\mathbf{k}'_0}(\mathbf{r}'' -
    \mathbf{r}')e^{i(\mathbf{k}'_0-\mathbf{k}_0) \mathbf{r}''}
    \label{pot_long}
\end{equation}
\end{widetext}
with
\begin{equation}
    \Delta_{\mathbf{k}_0}(\mathbf{r} - \mathbf{r}'')=\frac{1}{L^3} \sum\limits_{\mathbf{k}\in
    SBZ(\mathbf{k}_0)} e^{i\mathbf{k}(\mathbf{r}-\mathbf{r}'')},
    \label{delta}
\end{equation}
where $L^3$  is the crystal volume.

The functions $\Delta_{\mathbf{k}_0}(\mathbf{r} - \mathbf{r}'')$ and
$\Delta_{\mathbf{k}'_0}(\mathbf{r}'' - \mathbf{r}')$ are related to the geometrical properties of SBZ shown in Fig. 1. They have compact support in momentum space and are well-localized in position space acting like a low-pass filter function leading to a smoothing of the potential  $V(\mathbf{r}'')$.

The envelope function is smoothly varying over the region spanned by the functions   $\Delta_{\mathbf{k}_0}(\mathbf{r} - \mathbf{r}'')$ and
$\Delta_{\mathbf{k}'_0}(\mathbf{r}'' - \mathbf{r}')$. Here we make the approximation that the envelope function is almost constant within that region. When this is verified, the region in k-space occupied by the PW expansion of the envelope function is much smaller than the volume of SBZ. Consequently, the envelope function in (\ref{pot_long}) can be moved out of the integrals. The integration of $\Delta_{\mathbf{k}'_0}(\mathbf{r}'' - \mathbf{r}')$ over $\mathbf{r}'$ gives unity. The resulting equation reads:

\begin{equation}
    Vf=\sum\limits_{m,\mathbf{k}'_0} V_{\mathbf{k}_0,\mathbf{k}'_0}^{n,m}(\mathbf{r}) f_m(\mathbf{k}_0',\mathbf{r}),
\end{equation}
where:
\begin{eqnarray}
    V_{\mathbf{k}_0,\mathbf{k}'_0}^{n,m}(\mathbf{r})&&=\int d\mathbf{r}'' u_{n}^{*}(\mathbf{k}_0,\mathbf{r}'')V(\mathbf{r}'')u_{m}(\mathbf{k}_0',\mathbf{r}'')\nonumber \\
    &&\times \Delta_{\mathbf{k}_0}(\mathbf{r} - \mathbf{r}'')e^{i(\mathbf{k}'_0-\mathbf{k}_0)\mathbf{r}''}.
    \label{pot}
\end{eqnarray}

Eq. (\ref{pot}) has been derived assuming that the position of the impurity is fixed at the origin of the coordinate system. When it is not the case, it is easy to show that Eq. (\ref{pot}) has to be multiplied by a phase factor $e^{-i(\mathbf{k}'_0-\mathbf{k}_0)\mathbf{r}_0}$, where $\mathbf{r}_0$ is the position of the impurity atom. The phase factor becomes important for systems with more than one impurity atom \cite{us}.

The low-pass filtering of the product of the impurity potential and periodic Bloch functions in Eq. (\ref{pot}) is crucial, since this procedure eliminates nonphysical solutions which break the symmetry imposed by the crystal lattice. Such nonphysical envelope functions contain Fourier components with wave vectors $\mathbf{k}$ lying outside SBZ or outside BZ (see Fig. 1). The envelope functions with Fourier components lying outside corresponding SBZ, but inside BZ, lead to non-orthonormal wave functions $\psi(\mathbf{r})$. The envelope functions with Fourier components lying outside BZ are not consistent with periodic boundary conditions imposed by the crystal lattice. According to the Burt-Foreman definition of the envelope function, a fast-varying potential leads to band mixing keeping the envelope function smooth. Band mixing does not break the periodic boundary conditions of the crystal lattice. Here, we apply the single-band approximation and drop the band indices $n$  and $m$  in the potential (\ref{pot}) in the developments below.

The potential $V(\mathbf{r}'')$ in Eq. (\ref{pot}) may contain poles (e.g. the point charge potential), however singular points do not appear in the effective potentials $V_{\mathbf{k}_0,\mathbf{k}'_0}^{n,m}(\mathbf{r})$. When $\mathbf{k}_0 \neq \mathbf{k}'_0$, the singular point of the point charge potential always lies outside the compact support of the filter function, thus the integrand is not singular. The singularity may be found inside the compact support only when $\mathbf{k}_0 = \mathbf{k}'_0$. In this case, the principal value of the integral is computed.

\subsection{Effective potentials for P donors in silicon}

The effective potential defined by Eq. (\ref{pot}) can be treated as a non-local norm-conserving pseudo-potential since it appears in the effective mass equation. The non-locality is caused by the dependence of the potential on the wave vectors $\mathbf{k}_0$. Computed once for an impurity atom in bulk silicon, it can be used further, for example for silicon nanostructures with different confinement potentials.

For the phosphorus impurity, the bare potential $V(\mathbf{r}'')$ in Eq. (\ref{pot}) is modeled by the point charge Coulomb potential with the static dielectric screening \cite{Pantelides, Sven_potential}. This is in accordance with the Pantelides-Sah model \citep{Pantelides}, where the attractive potential is represented as a sum of two terms: $V=U_b+U_s$. The sum, $V$, is the residual between the silicon and silicon+impurity exact potentials. The term $U_b$ is the difference between silicon and phosphorus ionic potentials, the term $U_s$ is the difference in contributions coming from valence electrons. In the general case, the potential $U_b$ differs from the point charge potential by an effective charge which is not a constant and has a position-dependence. For isocoric impurities, this potential is very close to the constant 1.0, as proved by computations in Ref. [\onlinecite{Pantelides}] (see Fig. 4 within). Thus the point charge Coulomb potential is a good model in the case of phosphorus impurities. The contribution from $U_s$ is responsible for the static screening and can not be neglected. 
For the static screening we use the dielectric function computed by Nara \citep{Nara} from first principles using linear-response theory. However, the bare potential $V(\mathbf{r}'')$ of the non-isocoric impurities like As can not be modelled by the point charge potential. In this case, the bare potential can be computed using DFT or the simple ab-initio technique described in \citep{Pantelides} and inserted in Eq. (\ref{pot}).

The integral in Eq. (\ref{pot}) has been computed using the convolution theorem and the fast-Fourier transform algorithm implemented in MATLAB \cite{MATLAB}. The periodic Bloch functions in Eq. (\ref{pot}) have been computed in the framework of the density functional theory using the local density approximation and the projector-augmented wave method \cite{Torrent} (PAW) implemented in the ABINIT software \cite{Gonze}. We use the PAW method for silicon since it is able to reproduce all-electron wave-functions with an accurate charge density at nuclei \cite{hf}. The computations have been done in two steps: first a self- consistent computation with sparse grids in k-space is carried out to achieve fast convergence of the total energy. In the second step accurate non-self-consistent computations on the basis of the previous step are run for specific points in the reciprocal space. Convergence has been reached when the difference in total energy between cycles was less than $8.5\cdot 10^{-5}$ meV. As a numerical accuracy test, we have also used the periodic Bloch functions computed by the pseudopotential method. The comparison of the effective potentials for two sets of the periodic Bloch functions does not show any difference. This is a result of the filtering procedure which eliminates components with large wave vectors and gives the same results when the norm of basis functions is conserved within the unit cell. For the same reason, using pseudopotentials beyond LDA may improve the accuracy in the periodic Bloch function computations, but it does not affect much the effective potentials in the single-band approximation. Nevertheless, the accuracy of the periodic Bloch functions computation is crucial when one is interested in a local value of the wave function.

\begin{figure*}[t]
\centering
%\subfigure[]{\includegraphics[width=7.3cm]{ps1}}
\subfigure[]{\includegraphics[width=4.5cm]{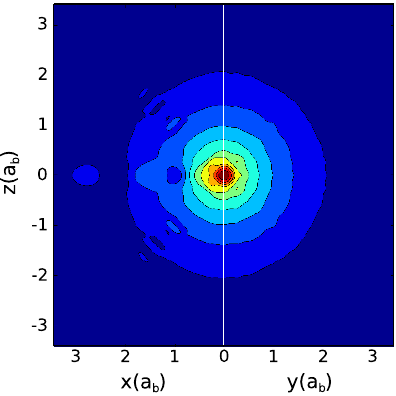}}
\subfigure[]{\includegraphics[width=4.5cm]{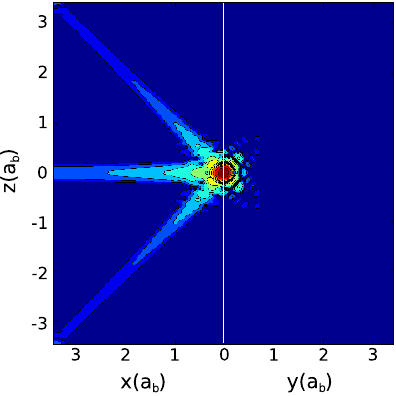}}
\subfigure[]{\includegraphics[width=4.5cm]{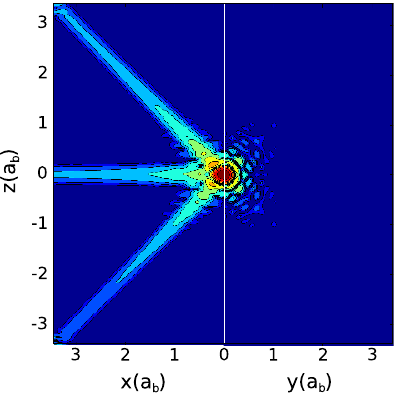}}
\caption{The effective potential $V_{\mathbf{k}_0,\mathbf{k}'_0}$ for a) $\mathbf{k}_0=\mathbf{k}'_0=X$, b) $\mathbf{k}_0=-\mathbf{k}'_0=X$ and c)
$\mathbf{k}_0=X$, $\mathbf{k}'_0=Y$  } 
\label{PS}
\end{figure*}

\begin{figure}[t]
\centering
\includegraphics[width=5.5cm]{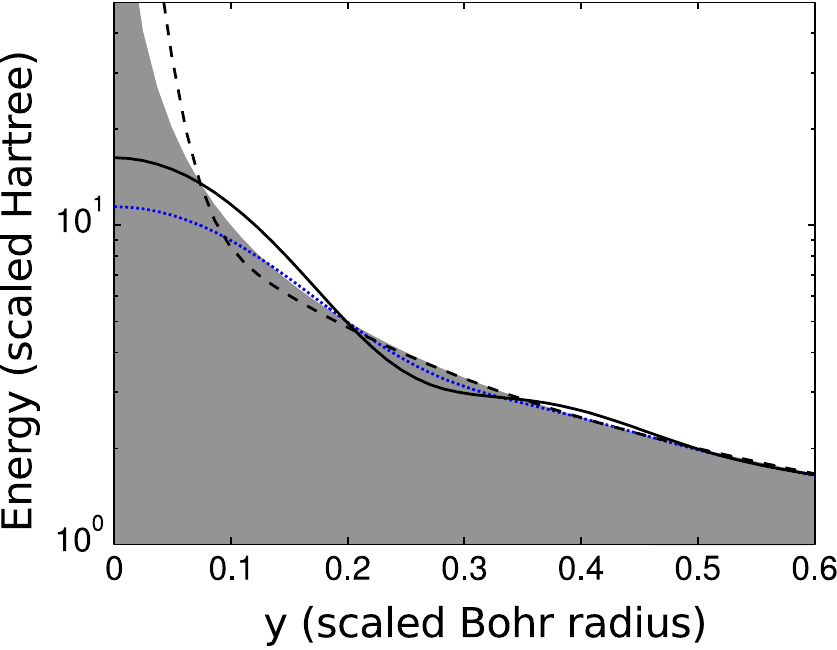}
\caption{Absolute value of the single-valley effective potential 
($\mathbf{k}_0=X$) plotted on a log scale, computed along the crystallographic axis [010]. The bare Coulomb potential and the screened Coulomb potential are designated by the shaded area and the black dashed line respectively. The corresponding effective potentials are represented by the blue dotted line (without screening) and the black solid line (with screening)} 
\label{P1}
\end{figure}

The computed effective potentials for a P donor atom in silicon are shown in  Fig. \ref{PS} for several combinations of valley indices $\mathbf{k}_0$ and $\mathbf{k}'_0$. The potential for the case when $\mathbf{k}_0=\mathbf{k}'_0$ (we call them the single-valley potentials) tends asymptotically to the Coulomb potential when  $|\mathbf{r}| \rightarrow \infty$. In Fig. \ref{P1} we compare the single-valley potentials with and without the static screening before and after applying the spatial filtering  (see Eq. (\ref{pot})). Unlike in the bare Coulomb potential, the central-cell region of the effective mass potential is smooth and does not contain singularity. When the static screening is neglected the single-valley effective-mass potential reaches its minimum at 11.62 scaled Hartrees (464.8 meV). Taking the screening into account decreases the potential energy minimum down to 16.46 scaled Hartrees (658.4 meV). The scaled units are defined in Appendix A. Also, the results of Fig. \ref{P1} show that the static screening affects the shape of the effective potential around the nucleus.

For the single-valley effective potential (Fig. \ref{PS} a), we observe Gibbs oscillations in the direction determined by the orientation of the constant energy ellipsoid associated with the conduction band valley. The oscillations are caused by boundaries of SBZ. They do not affect low-energy states localized around the core, however they have an effect on higher excited states. This effect is a consequence of the single-band approximation and it can be eliminated by including more bands.

The potentials for different wave vectors (we call them the coupling potentials) are more localized in real space and they are weaker. While the imaginary part of the single-valley potential is negligibly small, the real and imaginary parts of the coupling potentials are of the same order of magnitude.  Therefore, in  Fig. \ref{PS} b and c, we plot the absolute value. The coupling potentials possess a strong anisotropy. The coupling potential for which  $\mathbf{k}_0 \perp \mathbf{k}'_0$is approximately twice deeper comparing to that for which the condition $\mathbf{k}_0=-\mathbf{k}'_0$ holds.

Each effective potential separately does not exhibit the tetrahedral symmetry, but their proper combination does. In this sense, the effective potential is non-local, since in order to reproduce a proper symmetry the effective potential mixes envelope functions belonging to different valleys. Thus, the effective potential possesses a non-locality due to the dependence on the wave vectors $\mathbf{k}_0$ and $\mathbf{k}'_0$. The set of the effective potentials for all possible combinations of the wave-vectors can be arranged in a matrix which forms a reducible representation of the tetrahedral group. Therefore, the resulting wave function possesses the tetrahedral symmetry, whose significant importance has been discussed in Ref. \onlinecite{Gamble}.

\section{Solving  envelope function equations for single P donor in silicon}

\subsection{Numerical technique}

The resulting system of the envelope function equations consists in six coupled eigenvalue problems:

\begin{eqnarray}
    \left[H_{kp}(\mathbf{k}_0, \mathbf{k} \rightarrow i\nabla)
        +V_{\mathbf{k}_0,\mathbf{k}_0}(\mathbf{r})\right]f(\mathbf{k}_0,\mathbf{r})\nonumber \\
        +\sum\limits_{\mathbf{k}_0'\neq \mathbf{k}_0} V_{\mathbf{k}_0',\mathbf{k}_0}(\mathbf{r}) f(\mathbf{k}_0',\mathbf{r})=Ef(\mathbf{k}_0,\mathbf{r}).
\label{final_sys}
\end{eqnarray}

The envelope function equations  (\ref{final_sys}), written in the single-band approximation, are formally identical to the Shindo-Nara equations  \cite{Shindo-Nara,Gamble}. The difference with other approaches based on the Shindo-Nara equations lies in the definition of the potential  $V_{\mathbf{k}_0',\mathbf{k}_0}(\mathbf{r})$.

First we solve the problem neglecting valley-orbit coupling. Each single-valley equation with kinetic energy term $H_{kp}(\mathbf{k}_0,\mathbf{k} \rightarrow i\nabla)$ written explicitly reads:

\begin{eqnarray}
-\frac{1}{2}\left(\gamma_x \frac{\partial^2}{\partial
x^2}+\gamma_y\frac{\partial^2}{\partial y^2}  +\gamma_z
\frac{\partial^2}{\partial
z^2}\right)f^s(\mathbf{k}_0,\mathbf{r})\nonumber \\
+V_{\mathbf{k}_0,\mathbf{k}_0}(\mathbf{r})f^s(\mathbf{k}_0,\mathbf{r})
=E_{\mathbf{k}_0}^s f^s(\mathbf{k}_0,\mathbf{r}),
\label{hydros}
\end{eqnarray}
where $f^s(\mathbf{k}_0,\mathbf{r})$ and $E_{\mathbf{k}_0}^s$ are eigenfunctions and eigenvalues of the single-valley envelope function equations and $\gamma=\{\gamma_x,\gamma_y,\gamma_z\}=\{m_{xx}/m_{||}, m_{yy}/m_{||},m_{zz}/m_{||}\}$.
In bulk silicon, the components of the effective mass tensor $m_{xx}$ ,
$m_{yy}$ and $m_{zz}$ determine the orientation of the isoenergetic ellipsoids of each conduction band valley relative to the crystallographic axes. Thus, for different valleys the factors $\gamma$ take
different values: $\gamma=\{0.19,1,1\}$ for
valleys $\mathbf{k}_0=\{-X,X\}$, $\gamma=\{1,0.19,1\}$ for valleys
$\mathbf{k}_0=\{-Y,Y\}$ and $\gamma=\{1,1,0.19\}$
for valleys $\mathbf{k}_0=\{-Z,Z\}$.

The eigenfunctions and eigenvalues of Eq. (\ref{hydros}) are computed numerically using the finite element method with an unstructured grid adapted to the Coulomb potential of the impurity \cite{ffpp}.

Next to get the corrections caused by the central cell potential, we apply the variational method \cite{Morrison}, \cite{agoshkov}. First we expand the unknown envelope functions $f(\mathbf{k}_0,\mathbf{r})$ in terms of eigenfunctions
$f^s(\mathbf{k}_0,\mathbf{r})$ defined in Eq. (\ref{hydros}): $f(\mathbf{k}_0,\mathbf{r})=\sum\limits_{j,\mathbf{k}'_0}
c_{j,\mathbf{k}'_0} f^s_j(\mathbf{k}'_0,\mathbf{r})$, where $c_{j,\mathbf{k}'_0}$
is an expansion coefficient, and substitute this expansion in the system of equations (\ref{final_sys}). The eigenfunctions $f^s_j(\mathbf{k}'_0,\mathbf{r})$ taken from all valleys form a non-orthogonal basis set. Substituting the expansion in each equation of the system (\ref{final_sys}), multiplying by one of the basis functions and integrating over real space, one gets a system of linear algebraic equations:

\begin{equation}
\mathbf{B}\mathbf{C}=E\mathbf{S}\mathbf{C},
\end{equation}
where $\mathbf{C}$ is the vector of unknown expansion coefficients, $\mathbf{S}$ is the overlap matrix and $\mathbf{B}$ is a matrix with elements: 
\begin{equation}
    B_{\mathbf{k}_0,\mathbf{k}_0'}^{i,j}=\begin{cases}
    E_{j,\mathbf{k}_0}^s, & \text{if } \mathbf{k}_0=\mathbf{k}_0' \text{ and } i=j;\\
    M_{\mathbf{k}_0,\mathbf{k}_0'}^{i,j}, &\text{if } \mathbf{k}_0 \neq \mathbf{k}_0'
    \text{ or } i\neq j,
    \end{cases}
\end{equation}
where

\begin{equation}
    M_{\mathbf{k}_0,\mathbf{k}_0'}^{i,j}= \sqrt{6} \int d\mathbf{r}f^s_i(\mathbf{k}_0,\mathbf{r})V_{\mathbf{k}_0',\mathbf{k}_0}(\mathbf{r})
    f^s_j(\mathbf{k}'_0,\mathbf{r}).
    \label{nondiag_me}
\end{equation}

Eq. (\ref{nondiag_me}) can be further simplified taking into account the strong localization of the potential $V_{\mathbf{k}_0',\mathbf{k}_0}(\mathbf{r})$
(see discussion in the previous section) using proper asymptotic for atomic orbitals. The highest electron density at the nucleus is for s-type orbitals. Since the first term in their Tailor expansion is a constant, the matrix element can be rewritten as:

\begin{equation}
M_{\mathbf{k}_0,\mathbf{k}_0'}^{i,j}=\sqrt{6}f^s_i(\mathbf{k}_0,\mathbf{r}_0)f^s_j(\mathbf{k}'_0,\mathbf{r}_0) \int d\mathbf{r} V_{\mathbf{k}_0',\mathbf{k}_0}(\mathbf{r}).
    \label{nondiag_me_s}
\end{equation}

This approximation is identical to the contact potential approach \cite{Resca, Friesen, Friesen_well, Friesen_dot}.
 
\subsection{Binding energies}

The computed values of the three lowest energy levels of a P donor atom in silicon are collected in Table \ref{tab:energies}. The energies have been computed for two cases: for the bare Coulomb potential, $V_{bare}$, and for the potential with the static screening, $V_{scr}$ \cite{Pantelides}.

\begingroup
\squeezetable
\begin{table}[!h]
\caption{\label{tab:energies} Electron binding energies for P impurity in
silicon (meV)}
\begin{ruledtabular}
\begin{tabular}{lccc}
Symmetry                 & $A_1$ & $T_2$ & $E$ \\
\hline
$V_{bare}$ $(j=1s)$      &  -33.70  &  -32.65  & -32.58 \\
$V_{bare}$ $(j=1s,2s)$   &  -35.11  &  -34.22  & -34.16 \\
\hline
$V_{scr}$ $(j=1s)$       &  -43.23  &  -32.40  & -30.64 \\
$V_{scr}$ $(j=1s,2s)$    &  -45.40  &  -33.86  & -32.08 \\
\hline
Experiment \cite{Ramdas} &  -45.59  &  -33.89  & -32.58 \\
\end{tabular}
\end{ruledtabular}
\end{table}
\endgroup

The static screening screening leads to a small correction of 1.3 meV in the single-valley problem, however it affects more significantly the valley-orbit coupling potentials and leads to larger splitting energies. The best agreement with experimental data \cite{Ramdas} (within 0.5 meV) is obtained for the screened potential.

The method inherently takes into account the valley-orbit mixing between different single-valley orbitals. This kind of mixing has been first analyzed by Friesen  \citep{Friesen_dot} for quantum dots. Only orbitals of s-symmetry contribute to the VO mixing  because they have a large probability density at the nucleus, where the coupling potentials are localized  (see Fig. \ref{PS} a,b). From Table I, one can see that most significant contributions to the ground state energy come from the 1s orbital, while a non-negligible contribution of 0.5 meV is caused by 2s orbital.

We also compare in Table \ref{tab:me} the values of the matrix elements which are responsible for the valley-orbit coupling with those obtained in  Ref. [\onlinecite{Friesen}] from fitting to experimental data. We compare matrix elements computed for $j=1s$ only, since the fitting in Ref. [\onlinecite{Friesen}] has been done for 1s orbitals only.

\begingroup
\squeezetable
\begin{table}[!h]
\caption{\label{tab:me} Valley-orbit coupling matrix elements (meV)}
\centering
\begin{ruledtabular}
\begin{tabular}{lcccc}
Matrix element         & $E_0$   & $\Delta_0=E_0-E_H$ & $\Delta_1$ & $\Delta_2$ \\
\hline
Friesen \onlinecite{Friesen} & $E_H$=-31.28  & -4.13\footnotemark[1]  &  -1.51  & -2.17  \\
This work ($V_{bare}$) & -32.81  & -1.53  &  -0.12  & -0.17  \\
This work ($V_{scr}$)  & -33.70  & -2.42  &  -0.99  & -1.72 \\
\end{tabular}
\end{ruledtabular}
\footnotetext[1]{This number is computed as the difference between $E_H$ and experimental results.}
\end{table}
\endgroup

The matrix element $\Delta_0$ is the contribution from the central cell potential to the single-valley energy spectrum. This contribution is defined as the energy difference between  $E_H=-31.28$ meV, that is, the ground state energy from the hydrogenic model \citep{Friesen}, and the computed single-valley ground state energy, $E_0$. The matrix elements $\Delta_1$ and $\Delta_2$ are the values of the valley-orbit couplings defined by the integral in Eq. (\ref{nondiag_me_s}): $\Delta_j=\int d\mathbf{r} V_{\mathbf{k}_0',\mathbf{k}_0}(\mathbf{r})$, where $j=1$ for $\mathbf{k}_0=-\mathbf{k}'_0$ and $j=2$ for $\mathbf{k}_0 \perp \mathbf{k}'_0$. 

The difference between the present results and those of Friesen \citep{Friesen} is due to the fact that the fitting in Ref. [\onlinecite{Friesen}] is carried out for 1s orbitals only. For the effective potentials obtained in this work, using only 1s orbitals leads to an inaccuracy of several meV. So, a good agreement with experimental results is achieved by adding contribution from 2s orbitals. The overall shape of the wave function including 2s orbitals is slightly different from the shape of the wave function composed of 1s orbitals only.

\subsection{Wave functions and comparisons with STM images}

\begin{figure*}[t]
\centering
%\subfigure[]{\includegraphics[width=7.3cm]{ps1}}
\subfigure[]{\includegraphics[width=4.5cm]{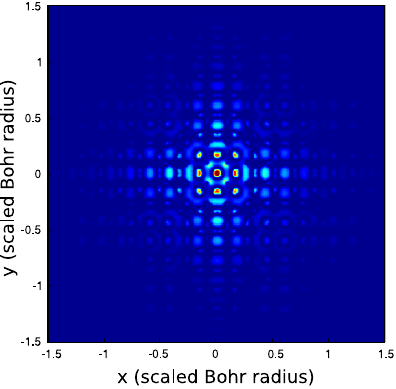}}
\subfigure[]{\includegraphics[width=4.5cm]{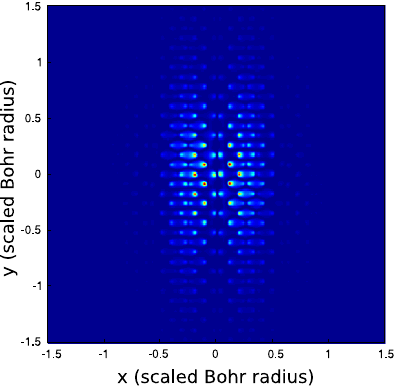}}
\subfigure[]{\includegraphics[width=4.5cm]{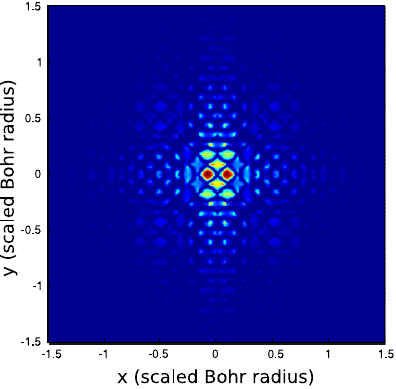}}
\caption{Wave functions of the P donor atom in silicon belonging to 1s manifold of states: a) non-degenerate $A_1$ state, b) triple-degenerate $T_2$ states, and c) double-degenerate $E$ states} 
\label{fig:wf}
\end{figure*}

In Fig. \ref{fig:wf} we report wave functions for the 1s manifold of states of the phosphorous donor atom. All wave functions have different symmetry within the unit cell, while their global shapes are similar and correspond to the contours of a 1s atomic orbital. The shape of the wave function shown in Fig. \ref{fig:wf} is in good agreement with results of Ref. [\onlinecite{Wellard}] and [\onlinecite{Abinitio}].

The computed value of $|\psi(\mathbf{r})|^2$ at the P nucleus for the ground state is $2.40 \times 10^{23}$ cm$^{-3}$, while the experimental value is $4.30 \times 10^{23}$ cm$^{-3}$ (see [\onlinecite{Kohn_book}] and [\onlinecite{Pica}] and references within). Recent results obtained using the supercell DFT computations with a small supercell  \cite{Overhof} show a little displacement of silicon atoms around the phosphorus impurity. Such short-range variations of the central-cell wave function can not be reproduced by adjustments made in the envelope function, but can be accounted for by including band mixing effect.

The wave function provides insights for interpreting the results of STM experiments. We compute here the wave function of the donor atom embedded below the silicon surface at 6.25 $a_0$, where $a_0$ is the lattice constant (see Appendix A). The surface electron density shown in Fig. \ref{val} is computed using the envelope-function approach to surface states described in Ref. [\onlinecite{Koenraad}] for 1x1 surface reconstruction. Also, we neglect valley-orbit splitting caused by the contact coupling at the surface \cite{Friesen_dot} since we consider the case when the donor atom is deep enough, so that the overlap of the wave function with the surface is negligibly small.

The results shown in Fig. \ref{val} are in semiquantitative agreement with STM measurements \citep{SvenVO}. The Fourier amplitudes recover the valley ellipsoids for -X, X, -Y and Y valleys positioned along the directions [100] and [010] in momentum space near the borders of the Brillouin zone. The figure also shows valley interference at the center of the Brillouin zone, which has been discussed in Ref. [\onlinecite{SvenVO}] in details. In Fig. 5 b, we plot the reciprocal space profile of the Fourier transform of the surface electron density in the direction [110]. The plot has three peaks: the central one defines the norm of the wave function (overall contribution from all valleys), while the side peaks indicates population of either from X- of from Y-valley. Therefore, by analysing these data it is possible to estimate valley population \citep{SvenVO}. Using analysis from Ref. [\onlinecite{SvenVO}], the population of Z-valley is estimated to be 43.9 \%, while the valley population in bulk silicon is the same for all valleys and equal to 33.33\%. Therefore, due to the effective mass anisotropy, the surface breaks valley degeneracy and leads to redistribution of the valley population enhancing the population of Z-valleys.

\begin{figure}[t]
\centering
\subfigure[]{\includegraphics[width=4cm]{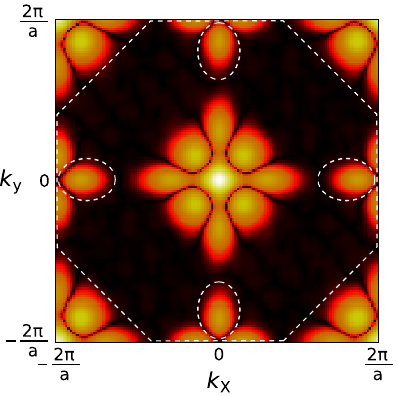}}
\subfigure[]{\includegraphics[width=4cm]{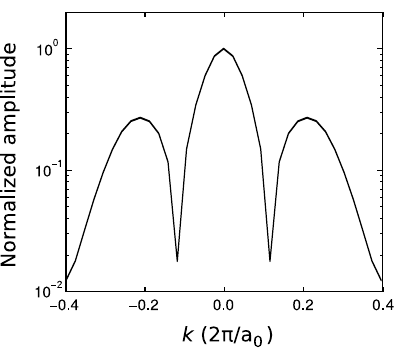}}
\caption{\label{val} a) Fourier amplitudes of the surface electron probability density $|\psi(\mathbf{r})|^2$ for the ground state computed by the envelope function approach for the P donor atom 6.25 $a_0$ below the silicon surface and b) the reciprocal space profile of $|\psi(\mathbf{r})|^2$ along the [110] direction.} 
\end{figure}

\section{Discussion and conclusions}

We derived envelope function equations based on the Burt-Foreman envelope function representation. The equations are free of fitting parameters and contain effective potentials describing electron confinement in the single-valley approximation, valley-orbit coupling and possibly interband coupling. The effective potentials are defined from the periodic Bloch functions and the exact confinement potential of a dopant through a low-pass filtering procedure which eliminates nonphysical Fourier components which are not consistent with the definition of the envelope functions, i.e., with the periodic boundary conditions imposed by the crystal lattice.

The potentials have been computed using \textit{ab initio} methods. The system of six envelope function equations has been solved in the single band approximation using the finite element method together with an eigenfunction expansion. The method proposed here takes into account valley-orbit coupling for different atomic orbitals. We have shown that the most significant contributions come from 1s and 2s atomic orbitals: taking into account valley-orbit coupling for 2s orbitals decreases the ground state energy by circa one meV. The static screening in silicon is essential: it results in 1 meV correction for the single-valley ground state energy and enhances the matrix elements describing valley-orbit coupling almost by one order of magnitude  (see Table \ref{tab:me}).

The results show a very good agreement (within 0.5 meV) with experimentally measured binding energies for all electronic states of the impurity atom (the maximal relative error is 1.53 \%). Such a good agreement confirms the validity of the single band approximation for computing the binding energies. The computed electron density at the phosphorus nucleus, which is more sensitive to atomistic details of the central cell, is smaller than the experimental value by the factor 1.79. For comparison, the value computed by the tight-binding method in Ref. [\onlinecite{Sven_potential}] is smaller than the experimental one by the factor 1.5. In the tight-binding computations, a single fitting parameter has been used to adjust the energy spectrum. The reasons for the inaccuracy in the computed value result from the single band approximation and from small displacements of silicon atoms relative to their positions in the periodic lattice. Small displacements lead to inaccuracies when periodic Bloch functions are used as a basis set for approximating the wave function in the central cell.

The computed results have been obtained using following approximations: the single-band approximation, the approximation that the phosphorus atom does not change positions of surrounding silicon atoms (by using periodic basis functions) and neglecting intrinsic non-locality of the potential energy term, which is equivalent to the contact potential approximation \cite{Resca, Friesen} (see Eq. \ref{nondiag_me_s}). The agreement can be further improved going beyond the single-band approximation by using, for example, the 2x2 kp-Hamiltonian proposed in Ref. [\onlinecite{Belyakov}].

In addition, we also modeled the electron density of the phosphorus donor atom embedded below the silicon surface and probed by the STM measurements \citep{SvenVO}. This observable is a sensitive test to the quality of the computed wave function at large distances from to the impurity nucleus. The comparison shows a good semiquantitative agreement: the valley population of the surface electron density is in a good agreement with the experimental data, while the features caused by the silicon surface reconstruction are not reproduced by the proposed method. 
The ability to reproduce the overall shape of the wave function of the impurity atom in silicon with a good agreement with experimental data opens the ways to accurately model electron-electron correlation effects in many-dopants many-electron systems.

\acknowledgments

This work was jointly supported by the proactive collaborative projects TOLOP (318397) and MULTI (317707) of the Seventh Framework Program of the European Commission. The authors acknowledge fruitful discussions with M. Verstraete, J. Bocquel, J. Salfi, and B. Voisin. F.R. acknowledges support from Fonds National de la Recherche Scientifique, Belgium, and S.R. acknowledges support from the the ARC DP scheme (DP120101825).

\appendix

\section{Silicon material parameters and scaled atomic units}

All material parameters used in computations are collected in
Table \ref{tab:si}. The lattice constant, relative permittivity and effective
masses have been taken from Ref. [\onlinecite{Wellard}] and Ref. [\onlinecite{Friesen}].

\begin{table}[!h]
\caption{\label{tab:si}Si material parameters}
\centering
\begin{tabular}{lcc}
\hline\hline
 Parameters             & Notation and units            & Values\\
\hline
 Lattice constant       & $a_0$ (\AA{})    & 5.43 \\
 Relative permittivity  & $\varepsilon$       & 11.4 \\
 Effective masses       & $m_{\bot}$          & 0.191 \\
                        & $m_{||}$            & 0.916 \\
 Conduction band minima &                     & \\
 wave number            & $|k_0|$ (nm$^{-1}$)   & 9.72 \\
\hline\hline
\end{tabular}
\end{table}

To study the electronic structure of donor atoms in silicon it is convenient to
use the system of scaled atomic units in order to simplify the formalism. The units are defined by the following formulae:

\begin{itemize}
\item The unit of length is the scaled Bohr radius defined by:
\begin{equation}
    a_b=\frac{4 \pi \hbar^2 \varepsilon \varepsilon_0}{m_{\perp} q^2}= 3.15
    \text{ nm},
\label{bohr}
\end{equation}
\item The energy is measured in the scaled Hartree:
\begin{equation}
    E_H=\frac{q^2}{4 \pi \varepsilon \varepsilon_0 a_b}= 40 \text{ meV}.
\label{hartree}
\end{equation}
\end{itemize}

\newpage
%\bibliographystyle{apsrev4-1}
%\bibliography{bib_silicon}{}


\end{document}